\documentstyle[psfig,emulateapj]{article}

\def\boldxi{\mbox{\boldmath $\xi$}}
\def\boldvarepsilon{\mbox{\boldmath $\varepsilon$}}
\def\bnabla{\mbox{\boldmath $\nabla$}}
\def\divxi{\bnabla\cdot{\boldxi}}
\def\etal{et al.\,}
\def\Bruntfreq{Brunt-V{\"a}is{\"a}l{\"a}\,\,\,}
\def\refnew#1{(\ref{#1})}
\def\be{\begin{equation}}
\def\ee{\end{equation}}

\def\K{\, \rm K}
\def\s{\, \rm s}

\def\cm{\, \rm cm}

\begin{document} 

\title{\mbox{GRAVITY-MODES IN ZZ CETI STARS} \\ \mbox IV.
AMPLITUDE SATURATION BY PARAMETRIC INSTABILITY}

\lefthead{Amplitudes of G-Modes in ZZ Cetis}
\righthead{Wu \& Goldreich}

\author{Yanqin Wu\altaffilmark{1,2}, Peter Goldreich\altaffilmark{3}}

\altaffiltext{1}{Astronomy Unit, School of Mathematical Sciences, 
	 Queen Mary and Westfield College, London, UK}
\altaffiltext{2}{Canadian Institute of Theoretical Astrophysics, 
	University of Toronto, 60 St. George Street, Toronto, Ontario M5S 3H8, Canada;
	wu@cita.utoronto.ca}
\altaffiltext{3}{Theoretical Astrophysics, California Institute of Technology 
	130-33, Pasadena, CA 91125, USA; pmg@gps.caltech.edu}

\begin{abstract}
ZZ Ceti stars exhibit small amplitude photometric pulsations in
multiple gravity modes. As the stars cool their dominant modes shift
to longer periods. We demonstrate that parametric instability limits
overstable modes to amplitudes similar to those observed. In
particular, it reproduces the trend that longer period modes have
larger amplitudes.

Parametric instability is a form of resonant 3-mode coupling. It
involves the destabilization of a pair of stable daughter modes by an
overstable parent mode. The 3-modes must satisfy exact angular
selection rules and approximate frequency resonance. The lowest
instability threshold for each parent mode is provided by the daughter
pair that minimizes $(\delta\omega^2+\gamma_d^2)/\kappa^2$, where
$\kappa$ is the nonlinear coupling constant, $\delta\omega$ is the
frequency mismatch, and $\gamma_d$ is the energy damping rate of the
daughter modes.  Parametric instability leads to a steady state if
$|\delta\omega|>\gamma_d$, and to limit cycles if
$|\delta\omega|<\gamma_d$. The former behavior characterizes low
radial order ($n\leq 3$) parent modes, and the latter those of higher
$n$.  In either case, the overstable mode's amplitude is maintained at
close to the instability threshold value.

Although parametric instability defines an upper envelope for the
amplitudes of overstable modes in ZZ Ceti stars, other nonlinear
mechanisms are required to account for the irregular distribution of
amplitudes of similar modes and the non-detection of modes with
periods longer than $1,200\s$. Resonant 3-mode interactions involving
more than one excited mode may account for the former. Our leading
candidate for the latter is Kelvin-Helmholtz instability of the
mode-driven shear layer below the convection zone.

\end{abstract}


\setcounter{equation}{0}

\section{INTRODUCTION}
\label{sec:nl-intro}

Within an instability strip of width $\Delta T_{\rm eff}\approx
10^3\K$ centered at $T_{\rm eff}\approx 1.2\times 10^3\K$, hydrogen
white dwarfs exhibit multiple excited gravity modes with $10^2\lesssim
P\lesssim 10^3\s$. Convective driving, originally proposed by
Brickhill (\cite{nonad-brick90}, \cite{nonad-brick91}), is the
overstability mechanism (Goldreich \& Wu \cite{nl-paperI}, hereafter
Paper I). Individual modes maintain small amplitudes; typical
fractional flux variations range from a few mma to a few tens of
mma.\footnote{$1$ mma of light variation is approximately $0.1\%$
fractional change in flux.}

The nonlinear mechanism responsible for saturating mode amplitudes has
not previously been identified. We demonstrate that parametric
resonance between an overstable parent g-mode and a pair of lower
frequency damped daughter g-modes sets an upper envelope to the parent
modes' amplitudes.\footnote{The g-mode dispersion relation allows
plenty of good resonances.}  Moreover, the envelope we calculate
reproduces the broad trends found from observational determinations of
mode amplitudes in ZZ Ceti stars. Our investigation follows pioneering
work by Dziembowski \& Krolikowska (\cite{nl-dziem85}) on overstable
acoustic modes in $\delta$-Scuti stars. They showed that parametric
resonance with damped daughter g-modes saturates the growth of the
overstable p-modes at approximately their observed amplitudes.

This paper is comprised of the following parts. In \S
\ref{sec:para-para} we introduce parametric instability for a pair of
damped daughter modes resonantly coupled to an overstable parent
mode. We evaluate the parent mode's threshold amplitude and describe
the evolution of the instability to finite amplitude.  \S 3 is devoted
to the choice of optimal daughter pairs. We discuss relevant
properties of 3-mode coupling coefficients, and the constraints
imposed by frequency resonance relations and angular selection
rules. Evaluation of the upper envelope for parent mode amplitudes set
by parametric resonance is the subject of \S 4. Numerical results are
interpreted in terms of analytic scaling relations and compared to
observations. \S 5 contains a discussion of a variety of issues
leftover from this investigation. Detailed derivations are relegated
to a series of Appendices.

The stellar models used in this investigation were provided by Bradley
(\cite{nl-bradley96}). Their essential characteristics are $M_{*} =
0.6 M_{\odot}$, $\log(g/\cm\s^{-2}) = 8.0$, hydrogen layer mass
$1.5\times 10^{-4} M_{*}$, and helium layer mass $1.5\times 10^{-2}
M_{*}$.

\section{PARAMETRIC INSTABILITY}
\label{sec:para-para}

In this section we introduce parametric instability (Landau \&
Lifshitz \cite{nl-landau76}). We present the threshold criterion for
the instability, and discuss relevant aspects of the subsequent
evolution. Depending upon the parameters, the modes either approach a
stable steady-state or develop limit cycles. We describe the energies
attained by the parent and daughter modes in either case. Many of the
results in this section were obtained earlier by Dziembowski
(\cite{nl-dziem82}).

\subsection{Instability Threshold}
\label{subsec:para-threshold}

Parametric instability in the context of our investigation is a
special form of resonant 3-mode coupling. It refers to the
destabilization of a pair of damped daughter modes by an overstable
parent mode. The frequencies of the three modes satisfy the
approximate resonance condition $\omega_p\approx
\omega_{d_1}+\omega_{d_2}$, where the subscripts $p$ and $d_1,d_2$
denote parent and two daughter modes, respectively.

Equations governing the temporal evolution of mode amplitudes are most
conveniently derived from an action principle (Newcomb \cite{nl-newcomb62}, 
Kumar \& Goldreich \cite{nl-kumar89}). The amplitude equations take the form
\begin{eqnarray}
{{d A_p}\over{dt}} &= &+{\gamma_p\over 2} A_p - i \omega_p A_p + i
{{3}\over{\sqrt{2}}} \omega_p\kappa A_{d_1} A_{d_2},
\label{eq:nl-amplieqnfulla}\\
{{d A_{d_1}}\over{dt}} &= &-{\gamma_{d_1}\over 2} A_2 - i \omega_{d_1} A_{d_1} + 
i
{{3}\over{\sqrt{2}}} \omega_{d_1}\kappa A_p A_{d_2}^*,
\label{eq:nl-amplieqnfullb}\\
{{d A_{d_2}}\over{dt}} &= &-{\gamma_{d_2}\over 2} A_{d_2} - i \omega_{d_2} 
A_{d_2} + i
{{3}\over{\sqrt{2}}} \omega_{d_2}\kappa A_p A_{d_1}^*.
\label{eq:nl-amplieqnfullc}
\end{eqnarray}
Here, $A_j$ is the complex amplitude of mode $j$; it is related to
the mode energy $E_j$ by $|A_j|^2 = E_j$. The $\gamma_j$ ($>0$) denote
linear energy growth and damping rates, and $\kappa$ is the nonlinear
coupling constant (cf. \S \ref{subsec:nl-kappa}). 
Our amplitude equations differ only in notation from those given by 
Dziembowski (\cite{nl-dziem82}).

The instability threshold follows from a straightforward linear
stability analysis applied to equations \refnew{eq:nl-amplieqnfullb}
and \refnew{eq:nl-amplieqnfullc}. The effects of nonlinear
interactions on the parent mode are ignored as is appropriate for
infinitesimal daughter mode amplitudes. The critical parent mode
amplitude satisfies (Vandakurov \cite{nl-vandakurov79}, Dziembowski
\cite{nl-dziem82}).  \be
|A_p|^2=\frac{\gamma_{d_1}\gamma_{d_2}}{18\kappa^2\omega_{d_1}\omega_{d_2}}\left
[1
+\left(\frac{2\delta\omega}{\gamma_{d_1}+\gamma_{d_2}}\right)^2\right],
\label{eq:thresh}
\ee
where $\delta\omega\equiv \omega_{d_1}+\omega_{d_2}-\omega_p$. 

\begin{figure*}[t]
\centerline{\psfig{figure=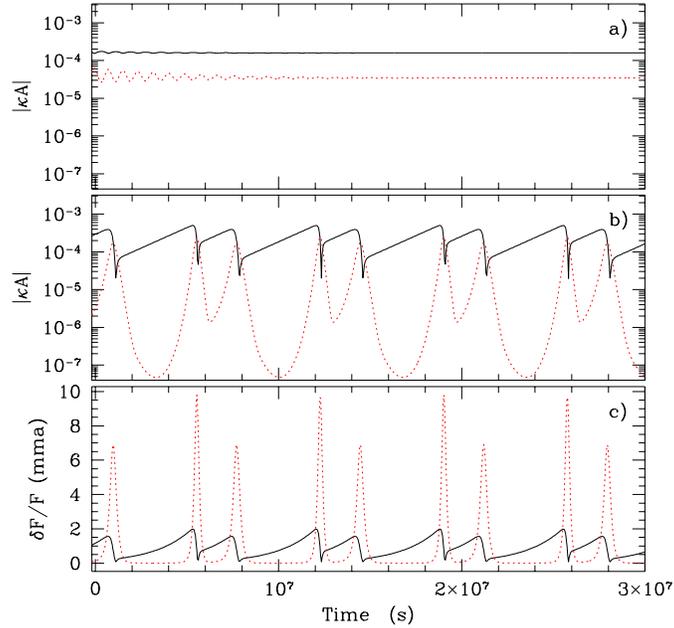,width=0.5\hsize}}
\caption[]{Parametric instability studied by numerical integration of
equations
\refnew{eq:nl-amplieqnfulla}-\refnew{eq:nl-amplieqnfullc}. Solid lines
represent parent mode amplitudes and dashed lines those of daughter
modes. We multiply these amplitudes with $\kappa$ to make them
dimensionless.  For the top panel $\delta \omega = \omega_{d_1} +
\omega_{d_2} - \omega_p = 2\times 10^{-5} \s^{-1} \gg\gamma_{d_1} =
\gamma_{d_2} = 10^{-6}\s^{-1}$, and the system settles into a steady
state.  The middle panel shows a case with $\delta \omega = 10^{-6}
\s^{-1}\ll \gamma_{d_1} = \gamma_{d_2} = 10^{-5} \s^{-1}$, for which
the mode amplitudes undergo limit cycles.  These panels illustrate the
two types of behavior discussed in \S \ref{subsec:para-time}. The
bottom panel displays photospheric flux variations associated with the
case shown in the middle panel. The daughter mode energies
episodically approach that of the parent mode. Around these times
their fractional flux amplitudes exceed that of the parent mode
because they have smaller mode masses. For both simulations, we take
mode periods to be $P_p = 500 \s^{-1}$, $P_{d_2}/P_{d_1}=0.9$, and
$\gamma_p = 10^{-7}\s^{-1}$.}
\label{fig:nl-detail} 
\end{figure*}

\begin{figure*}[t] 
\centerline{\psfig{figure=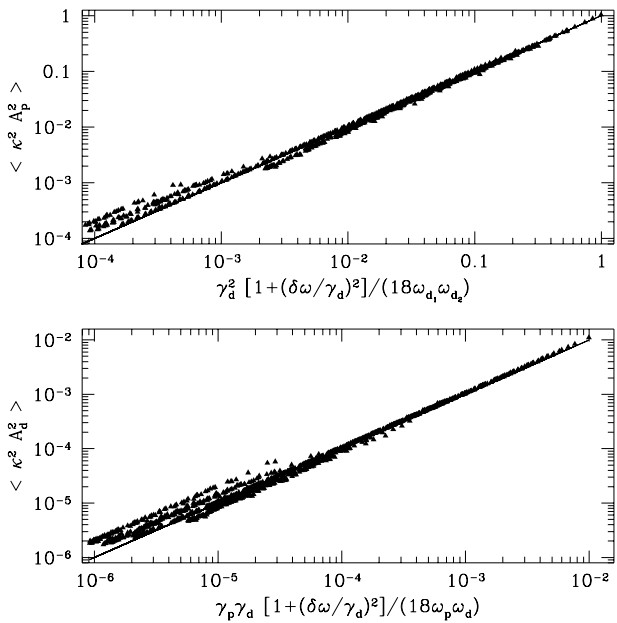,width=0.5\hsize}} 
\caption[]{Numerical simulations confirm theoretical predictions for
average mode energies.  We numerically integrate equations
\refnew{eq:nl-amplieqnfulla}-\refnew{eq:nl-amplieqnfullc} for a
variety of three-mode systems with different frequencies and energy
growth/decay rates. Each point on the figure represents one such
system. For simplicity, we restrict ourselves to systems for which
$\gamma_{d_1} = \gamma_{d_2} = \gamma_d$.  Time-averaged energies of
parent and daughter modes are plotted against theoretically predicted
values (eqs. [\ref{eq:nl-amplisteady1}]-[\ref{eq:nl-amplisteady3}]) in
the upper and lower panels, respectively. Energies multiplied by
$\kappa^2$ are dimensionless. The subscript `d' in the lower panel is
taken to be $d_1$ or $d_2$.}
\label{fig:nl-formula} \end{figure*}

\subsection{Dynamics}
\label{subsec:para-time}

The amplitude equations 
\refnew{eq:nl-amplieqnfulla}-\refnew{eq:nl-amplieqnfullc} have a
unique equilibrium solution given by
\begin{eqnarray}
|A_p|^2 & = &
\frac{\gamma_{d_1}\gamma_{d_2}}{18\kappa^2\omega_{d_1}\omega_{d_2}}\left[1
+\left(\frac{2\delta\omega}{\gamma_{d_1}+\gamma_{d_2}-\gamma_p}\right)^2\right],
\label{eq:nl-amplisteady1}\\
|A_{d_1}|^2 & = & 
\frac{\gamma_{d_2}\gamma_p}{18\kappa^2\omega_{d_2}\omega_p}\left[1
+\left(\frac{2\delta\omega}{\gamma_{d_1}+\gamma_{d_2}-\gamma_p}\right)^2\right],
\label{eq:nl-amplisteady2}\\
|A_{d_2}|^2 & = & 
\frac{\gamma_{d_1}\gamma_p}{18\kappa^2\omega_{d_1}\omega_p}\left[1
+\left(\frac{2\delta\omega}{\gamma_{d_1}+\gamma_{d_2}-\gamma_p}\right)^2\right],
\label{eq:nl-amplisteady3}
\end{eqnarray}
together with
\begin{equation}
\cot \Phi = - {{2 \delta \omega}\over{\gamma_{d_1} + \gamma_{d_2} - \gamma_p}}.
\label{eq:nl-amplisteadyphi}
\end{equation}
Here, $\Phi = \theta_{d_1} + \theta_{d_2} - \theta_p$, where the
complex amplitude $A_j$ may be written as $A_j = |A_j| e^{-i
\theta_j}$.  Note that equation \refnew{eq:nl-amplisteady1} is almost
identical to the threshold criterion (eq. [\ref{eq:thresh}]) for
$\gamma_p \ll \gamma_d$ which is the case that concerns us. Here
$\gamma_d = (\gamma_{d_1} +\gamma_{d_2})/2$ is the characteristic
damping rate for the daughter modes. It is also worth mentioning that,
in this limit, the parent mode energy is independent of $\gamma_p$.

The equilibrium state is a stable attractor for mode triplets with
$|\delta \omega| > \gamma_d$, and unstable otherwise (Wersinger \etal
\cite{nl-wersinger80}, Dziembowski \cite{nl-dziem82}). Figures
\ref{fig:nl-detail}a \& \ref{fig:nl-detail}b illustrate these two
types of behaviors. Triplets with unstable equilibria undergo a
variety of limit cycles. These share a number of common features. The
parent mode's amplitude remains close to its equilibrium (threshold)
value, with slow rises on time scale $\gamma^{-1}_p$ followed by
precipitous drops on time scale $\gamma^{-1}_d$. The daughter modes'
amplitudes stay far below their equilibrium values for most of the
cycle, but peak with amplitudes comparable to that of the parent mode
for a brief interval of length $\gamma^{-1}_d$ shortly after the
parent mode amplitude reaches its maximum value. During this brief
interval, the energy which the parent mode has slowly accumulated is
transferred to and dissipated by the daughter modes. So we see that,
independent of its stability, the equilibrium state defines the parent
mode's amplitude. This is confirmed by Figure
\ref{fig:nl-formula}a. We employ this result in \S \ref{sec:nl-gmode}
where we predict upper limits for g-mode amplitudes in pulsating white
dwarfs. The time-averaged energies of the daughter modes are also
found to be consistent with equations
\refnew{eq:nl-amplisteady2}-\refnew{eq:nl-amplisteady3}
(Fig. \ref{fig:nl-formula}b).

\section{CHOOSING THE BEST DAUGHTER PAIRS}
\label{sec:nl-gmode}

The discussion in the previous section shows that an overstable parent
mode's amplitude saturates at a value close to the threshold for
parametric instability.  Although each overstable parent mode has many
potential daughter pairs, the most important pair is the one with the
lowest instability threshold. This section is devoted to identifying
these optimal daughter pairs, a task which separates into two
independent parts, maximization of $\kappa^2$ and minimization of
$(\delta \omega^2 + \gamma_d^2)$.

\subsection{Three-Mode Coupling Coefficients}
\label{subsec:nl-kappa}

The 3-mode coupling coefficient characterizes the lowest order
nonlinear interactions among stellar modes. A compact form suitable
for adiabatic modes under the Cowling approximation is derived in 
Kumar \& Goldreich (\cite{nl-kumar89});
\begin{eqnarray}
\kappa & = & - \int d^3 x {{p}\over 6} \left\{
 (\Gamma_1 -1)^2 (\divxi)^3 \right. \nonumber  \\
& & \left. + 3 (\Gamma_1 -1) (\divxi) \xi^i_{;j} \xi^j_{;i} 
+ 2 \xi^i_{;j} \xi^j_{;k} \xi^k_{;i} 
\right\},
\label{eq:app-kappaoriginal}
\end{eqnarray}
where $p$ is the unperturbed pressure, $\Gamma_1$ is the adiabatic
index, $\boldxi$ is the Lagrangian displacement, the symbol `;'
denotes covariant derivative, and the integration is over the volume of
the star. This expression for $\kappa$ is symmetric with respect to
the three modes. Note that the displacements enter only through
components of their gradients. The present form is not suitable for
accurate numerical computation. We derive a more appropriate version
in \S \ref{subsec:app-long} (eq. [\ref{eq:working2}]).

\begin{figure*}[t]
\centerline{\psfig{figure=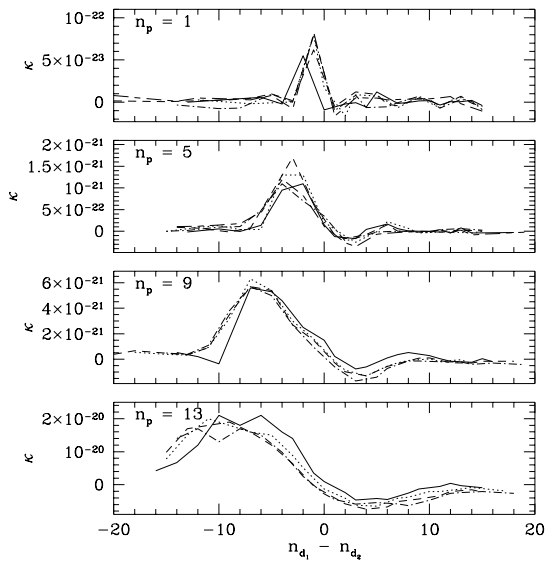,width=0.44\hsize}}
\caption[]{Coupling coefficient $\kappa$ in ${\rm erg}^{-1/2}$ for a
DA white dwarf model with $T_{\rm eff} = 12,800 \K$. From top to
bottom panels, we consider four ${\ell_p} = 1$ parent modes with
increasing $n_p$. Coupling coefficients between each parent and its
daughter pairs are plotted versus $n_{d_1} - n_{d_2}$. Solid lines
denote daughter pairs with ${\ell_{d_1}} = 1$, and others those with
${\ell_{d_1}} = 2,3,4,5$, and $9$. Mode $3$ is chosen to minimize
$|\delta \omega|$ subject to the constraint ${\ell}_{d_2} = {\ell}_p +
{\ell}_{d_1}$. $\kappa$ is seen to depend on the radial similarity of
the two daughter modes (eq. [\ref{eq:ampli-n2n3relation}]) but not on
their spherical degrees. The FWHM of the peak in $|\kappa|$ is of
order $n_p$. Maximum $|\kappa|$ occurs for $n_{d_1} - n_{d_2} \sim
-0.7 n_p$ (eq. [\ref{eq:dependn2n3}]).}
\label{fig:ampli-dependn2n3} 
\end{figure*}

Each eigenmode of a spherical star is characterized by three
eigenvalues $n,\ell,m$; $n$ is the number of radial nodes in the
radial displacement eigenfunction, $\ell$ is the spherical degree, and
$m$ is the azimuthal number.\footnote{The angular dependence is
described by a spherical harmonic $Y_{lm} (\theta,\phi)$.}
Integration over solid angle enforces the following selection rules on
triplets with non-vanishing $\kappa$: $|{\ell}_{d_2}-{\ell}_{d_1}|\leq
{\ell}_p \leq {\ell}_{d_1}+{\ell}_{d_2}$, ${\ell}_p+{\ell}_{d_1}
+{\ell}_{d_2}$ even, and $m_p = m_{d_1} + m_{d_2}$ (see \S
\ref{subsec:app-long}). These selection rules guarantee the
conservation of angular momentum during nonlinear interactions.  The
magnitude of $\kappa$ is largest when the eigenfunctions of the
daughter modes are radially similar in the upper evanescent zone of
the parent mode.  Radial similarity requires near equality of the
vertical components of WKB wavevectors, $k_z$, where for gravity modes
$k_z^2 \approx (N^2/\omega^2 - 1) \Lambda^2/r^2$ (eq. [A2] of Paper I)
with $N^2$ being the \Bruntfreq frequency, $r$ the radius, and
$\Lambda^2 = \ell(\ell + 1)$. We take $N \sim \omega_p$ as the major
contribution to the peak value of $|\kappa|$ comes from the region
just above $z_{\omega_p}$, the upper boundary of the parent mode's
propagating cavity. The g-mode dispersion relation applicable for
$\omega \ll 10^{-2} \s^{-1}$ is $\omega \propto \ell/n$ (see Fig. 4 of
Paper I).  We find that $\kappa$ is largest when
\begin{equation}
{{n_{d_1}}\over{n_{d_2}}} \sim {{\Lambda_{d_1} \omega_{d_1}}\over{\Lambda_{d_2} 
\omega_{d_2}}}
\sim \left({{\omega_p^2 - \omega_{d_1}^2}\over{\omega_p^2 - 
\omega_{d_2}^2}}\right)^{1/2}.
\label{eq:ampli-n2n3relation}
\end{equation}
Taking $\ell_p = 1$, $\ell_{d_2} = \ell_{d_1} +1$, and evaluating the
above relation for $\ell_{d_1} < 10$, we locate the peak of $\kappa$
to be at
\begin{equation}
n_{d_1}-n_{d_2}\approx -0.7 n_p . 
\label{eq:dependn2n3}
\end{equation}
Since a fraction $n_p^{-1}$ of the nodes of each daughter mode lie in
the region above $z_{\omega_p}$, the peak has a width of order $n_p$ 
when measured in $n_{d_1}-n_{d_2}$. Figure \ref{fig:ampli-dependn2n3} 
illustrates the behavior of $\kappa$ for a variety of combinations of parent and
daughter modes.

\begin{figure*}[t]
\centerline{\psfig{figure=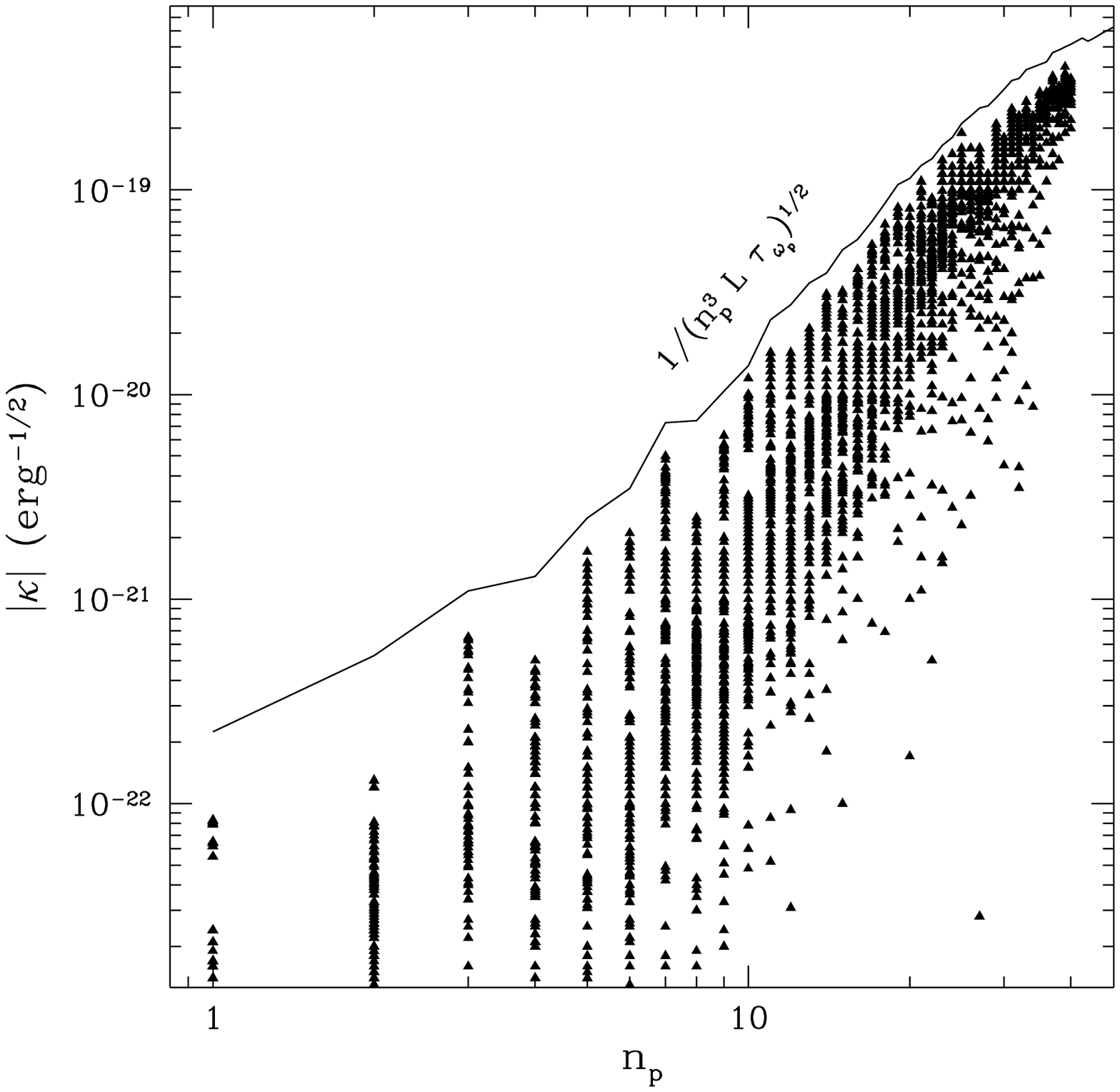,width=0.43\hsize}}
\caption[]{Coupling coefficients as a function of parent mode's radial
order.  We consider $\ell_p = 1$ parent modes, and include all
daughter pairs that have ${\ell}_{d_1} = {1,...,9}$ and ${\ell}_{d_2}
= {\ell}_p + {\ell}_{d_1}$. For each mode $2$, mode $3$ is chosen to
minimize $|\delta\omega|$.  The solid line corresponds to the
theoretical estimate for the maximum $|\kappa|$ at each $n_p$
(eq. [\ref{eq:nl-kappamax}]). $L$ is the stellar luminosity, and
$\tau_{\omega_p}$ is the thermal time at $z_{\omega_p}$. These
calculations are based on the same stellar model used for Figure
\ref{fig:ampli-dependn2n3}.}
\label{fig:ampli-kappaeulerian} 
\end{figure*}

As we show in \S \ref{subsec:app-orders}, the maximum value of $|\kappa|$ is of 
order
\begin{equation}
|\kappa|_{\rm max} \sim \frac{1}{\left(n_p^3 \tau_{\omega_p} L\right)^{1/2}}.
\label{eq:nl-kappamax}
\end{equation}
Here $L$ is the stellar luminosity and $\tau_{\omega_p}$ is the
thermal timescale at $z_{\omega_p}$. This is compared with numerical
results in Figure \ref{fig:ampli-kappaeulerian}. Note that the maximum
value of $\kappa$ depends entirely upon the properties of the parent
mode and not at all upon those of its daughters.

\subsection{Frequency Mismatch and Damping Rates}
\label{subsec:nl-others}

Because the maximum value of $|\kappa|$ is independent of $\ell_{d_1}$ and
$\ell_{d_2}$, we choose these parameters to minimize
$(\delta \omega^2 + \gamma_d^2)$.

Consider an $\ell_p, m_p$ parent mode. For each choice of $\ell_{d_1},
\ell_{d_2}$, there are of order $\ell_{d_1} n_p^2$ daughter pairs for
which $\kappa$ is close to its maximum value. The factor $\ell_{d_1}$
arises from the freedom in choosing $m_{d_1}$, while the factor
$n_p^2$ comes from the width of maximum $|\kappa|$ at each
$\ell_{d_1}$. Relaxing the value of $\ell_{d_2}$ subject to the
constraint of the angular selection rules increases the number of
pairs by a factor of order $\ell_p$. Now replace $\ell_{d_1}$ by a
running variable ${\ell_{d_1}}^\prime$. The number of pairs with
${\ell_{d_1}}^\prime \leq \ell_{d_1}$ is of order $\ell_p \ell_{d_1}^2
n_p^2$. The distribution of the $\delta \omega$ values of these pairs
is uniform between $0$ and $n_p\omega_{d_1}/n_{d_1}$.\footnote{The factor
$n_p$ arises because the peak in $\kappa$ has a width
$|n_{d_1}-n_{d_2}|\sim n_p$.}  Statistically, the minimum frequency
mismatch
\begin{equation}
\Delta \omega \sim {{\omega_{d_1}}\over{n_{d_1}}}{1\over{\ell_p \ell_{d_1}^2 
n_p}}
\sim {{\omega_p}\over{\ell_{d_1}^3 n_p^2}},
\label{eq:app-deltaomega}
\end{equation}
where the low frequency limit of the dispersion relation,
$\omega\propto \ell/n$, is assumed in going from the first to the
second relation for $\Delta\omega$. Our estimate for $\Delta\omega$
assumes that rotation lifts $m$ degeneracy. If it does not, the
minimum frequency mismatch is increased by a factor of $\ell_{d_1}$.

\begin{figure*}[t]
\centerline{\psfig{figure=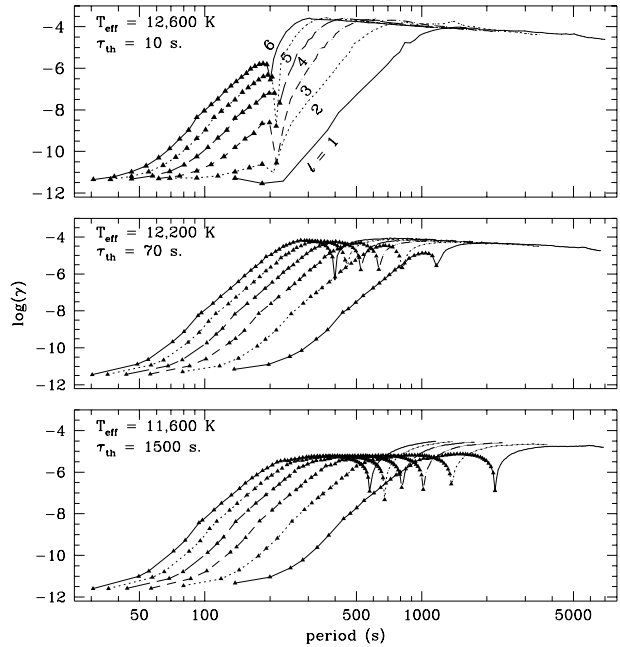,width=0.5\hsize}}
\caption[]{Nonadiabatic energy driving and damping rates as a function
of mode period. See Paper II for computational details.  We consider
three stellar models with decreasing effective temperature; $\tau_{\rm
th}$ is the thermal time scale at the bottom of the convective zone.
Overstable modes are marked by solid triangles. Lines connect points
associated with modes of the same $\ell$. For each $\ell$, $\gamma$
exhibits an initial steep rise with increasing mode period
(eq. [\ref{eq:fastgamma}]), followed by a gradual decline
(eq. [\ref{eq:slowgamma}]). Maximum $\gamma$ occurs at the
transition between quasi-adiabatic and strongly nonadiabatic damping.}
\label{fig:ampli-omegai}
\end{figure*}

Next we describe how $\gamma$ varies with $\omega$ and $\ell$. 
There are two regimes of relevance to this investigation. In the quasiadiabatic 
limit (cf. \S 4.4 of paper I),\footnote{Overstable modes are quasiadiabatic so 
this estimate for $\gamma$ applies to them as well as to damped modes.} 
\begin{equation}
\gamma \sim {1\over{n \tau_{\omega}}} \propto \left({\ell\over 
\omega}\right)^{6}.
\label{eq:fastgamma}
\end{equation} 
In the strongly nonadiabatic limit (see \S
\ref{subsec:app-gamma}),
\begin{equation}
\gamma \sim {\omega \over{\pi n}} \ln {1\over{\cal R}}\propto
\omega^{0.75}\ell^{0.2},
\label{eq:slowgamma}
\end{equation}
where ${\cal R}$ is the amplitude reflection coefficient at the
top of the mode's cavity. The transition between the quasiadiabatic
and strongly nonadiabatic limits is marked by a significant reduction
of ${\cal R}$ by radiative diffusion. The behavior of $\gamma$ as a
function of $\omega$ and $\ell$ is illustrated in Figure
\ref{fig:ampli-omegai}. Radial similarity of the daughter modes to
which a given parent mode couples most strongly
(eq.[\ref{eq:ampli-n2n3relation}]) implies $\gamma_{d_1}\approx
\gamma_{d_2}\approx \gamma_d$. 

\begin{figure*}[t]
\centerline{\psfig{figure=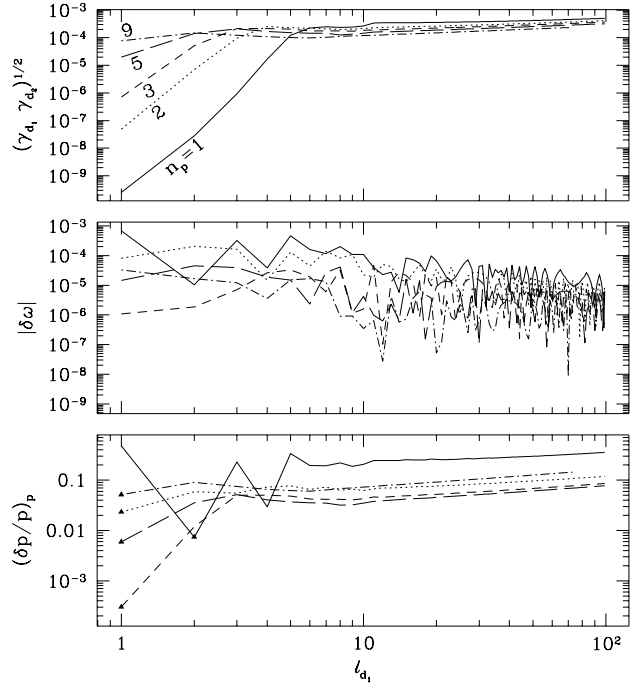,width=0.5\hsize}}
\caption[]{Selection of the best daughter pairs for a few low order
$\ell_p = 1$ parent modes in a white dwarf model with $T_{\rm eff} =
12,800 \K$. We look at daughter pairs with $\ell_{d_1}$ ranging from
$1$ to $100$, and $\ell_{d_2} = \ell_{d_1} +\ell_p$. We plot
$(\gamma_{d_1} \gamma_{d_2})^{1/2}$ in the upper panel, and $|\delta
\omega|$ in the middle panel, for the pair that sets the lowest
instability threshold at each $\ell_{d_1}$. This does not necessarily
correspond to the pair that give rise to the lowest $|\delta \omega|$
at each $\ell_{d_1}$. The corresponding threshold value of $(\delta
p/p)_p$ evaluated at the stellar surface is shown in the bottom panel.
With increasing $\ell_{d_1}$, $(\gamma_{d_1} \gamma_{d_2})^{1/2}$
rises and $|\delta\omega|$ becomes unimportant. The best daughter
pairs have low $\ell$ values.}
\label{fig:l2-dependence}
\end{figure*}

The minimum of $(\delta \omega^2 + \gamma_d^2)$ is attained for
daughter pairs that satisfy $\delta \omega^2 \lesssim \gamma_d^2$.
With increasing $n_p$, the value of $\ell_{d_1}$ at which this occurs
decreases from a few to unity (see Fig. \ref{fig:l2-dependence}). For
sufficiently large values of $n_p$, $\Delta\omega\ll \gamma_d$ even
for $\ell_{d_1}=1$.

\section{UPPER ENVELOPE OF PARENT MODE AMPLITUDES}
\label{sec:nl-estbest}

Parametric instability provides an upper envelope to the amplitudes 
of overstable modes. Coupling of an overstable parent mode to a single pair of 
daughter modes suffices to maintain the parent mode's amplitude near the
threshold value (cf. \S \ref{subsec:para-time}). The energy gained by the 
overstable parent mode and transferred to the daughter modes may be disposed of 
by linear radiative damping or by further nonlinear coupling to granddaughter 
modes.

The results presented in this section are obtained from numerical computations 
and displayed in a series of figures. The general trends they exhibit are best 
understood in terms of analytic scaling relations. We derive these first in 
order to be able to refer to them as we describe each figure. 

Excited modes of ZZ Ceti stars are usually detected through photometric 
measurements of flux variations, and in a few cases through spectroscopic 
measurements of horizontal velocity variations. Hence we calculate surface 
amplitudes of fractional flux, $\Delta F/F$, and horizontal velocity, $v_h$, 
variations.\footnote{In this section we drop subscripts on parent mode 
parameters and denote daughter mode parameters by a subscript $d$.}  
Each of these is directly related to the near surface amplitude of 
the Lagrangian pressure perturbation, $\delta p/p$. The threshold value of the 
latter is obtained by combining equations \refnew{eq:thresh} and 
\refnew{eq:nl-kappamax} with the normalization factor $(n\tau_\omega L)^{-1/2}$ 
given by equation (A28) of Paper 
I. Thus
\be
{\delta p\over p}\sim n\left[\left({\gamma_d\over 
\omega}\right)^2+\left({\delta\omega\over \omega}\right)^2\right]^{1/2}.
\label{eq:delp}
\ee
Adopting relations expressing $v_h$ and $\Delta F/F$ in terms of $\delta p/p$ 
from \S 3 of Paper I, we arrive at
\be
v_h\approx {\omega R n\over [\ell(\ell+1)]^{1/2}}
\left[\left({\gamma_d\over \omega}\right)^2+\left({\delta\omega\over 
\omega}\right)^2\right]^{1/2},
\label{eq:vhth}
\ee
and
\be
{\Delta F\over F}\approx {n\over 
\left[1+\left(\omega\tau_c\right)^2\right]^{1/2}}
\left[\left({\gamma_d\over \omega}\right)^2+\left({\delta\omega\over 
\omega}\right)^2\right]^{1/2},
\label{eq:DelFth}
\ee
In the above, $R$ denotes the stellar radius, and $\tau_c$ is the thermal time 
constant describing the low pass filtering action of the convection zone
on flux variations input at its base.\footnote{$\tau_c\approx 3\tau_{\rm th}$, 
where $\tau_{\rm th}$ is evaluated at $z_b$.}

\begin{figure*}[t]
\centerline{\psfig{figure=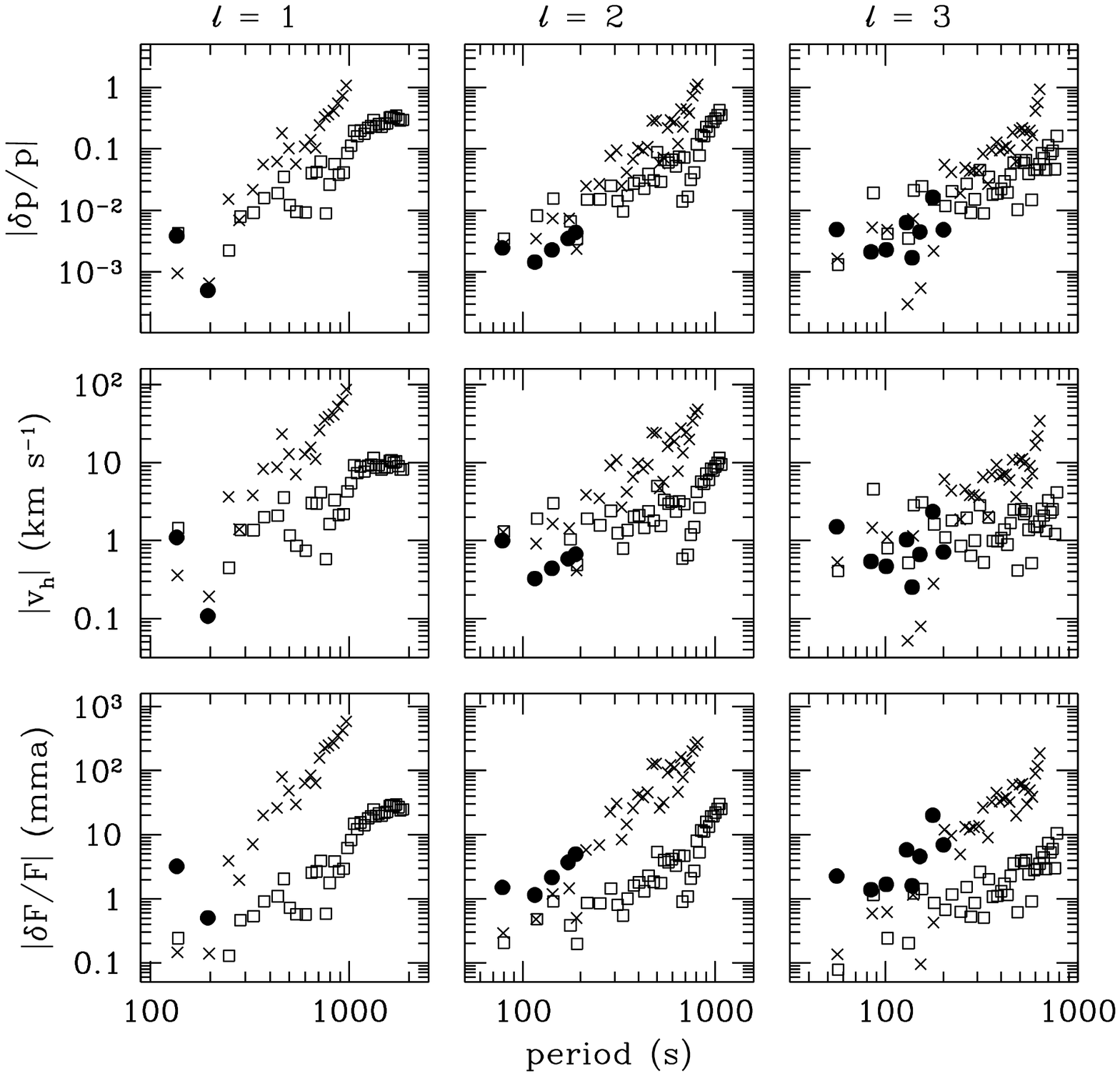,width=0.55\hsize}}
\caption[]{An assembly of results on threshold amplitudes for
parametric instability. From left to right, amplitudes of overstable
modes with different $\ell$ are plotted against mode period. From top
to bottom, the panels display photospheric amplitudes of fractional
pressure perturbation, horizontal velocity, and fractional flux
variation. Filled circles, crosses, and open squares symbolize modes
from the three stellar models in Figure \ref{fig:ampli-omegai} in
order of descending $T_{\rm eff}$. Pulsation amplitudes generally rise
with increasing mode period, except for the shortest period modes
which exhibit large star to star fluctuations. Amplitudes of $|\delta
p/p|$ for $\ell = 1$ modes in the coolest model dip around $600 \s$
and flatten beyond $1200 \s$. The first feature is associated with the
dip in $\gamma$ for neutrally stable daughter modes, and the second is
explained by the flattening of $\gamma$ for strongly nonadiabatic
daughter modes of long period (see Fig. \ref{fig:ampli-omegai}).
Reduction of the coupling coefficient due to strong nonadiabaticity is
not accounted for (see appendix \ref{subsec:app-nonad}).}
\label{fig:ampli-final}
\end{figure*}

Figure \ref{fig:ampli-final} displays calculated values for amplitudes of 
overstable modes limited by parametric instability. The rise of $|\delta p/p|$ 
with increasing mode period mainly reflects the corresponding rise of the 
damping rates of the daughter modes (cf. \S \ref{subsec:nl-others})
as indicated by equation \refnew{eq:delp}. At the longer periods, the values of 
$|\delta p/p|$ decline with decreasing $T_{\rm eff}$. This is a subtle 
consequence of the deepening of the convection zone which pushes down the top of 
the daughter modes' cavities, thus reducing their damping rates (see \S 5.2 of 
Wu \& Goldreich \cite{nl-paperII}, hereafter Paper II). The behavior of $|v_h|$ 
is similar to that of $|\delta p/p|$ except 
that $|v_h|$ decreases relative to $|\delta p/p|$ with increasing $\ell$ as 
shown by comparison of equations \refnew{eq:delp} and \refnew{eq:vhth}. A new 
feature present in the run of $|\delta F/F|$ verses mode period is the low pass 
filtering action of the convection zone as expressed by the factor 
$[1+(\omega\tau_c)^2]^{-1/2}$ in equation \refnew{eq:DelFth}. This factor causes 
$|\delta F/F|$ to rise slightly more steeply than $|\delta p/p|$ with increasing 
mode period. It is also responsible for a more dramatic decrease in $|\delta 
F/F|$ with decreasing $T_{\rm eff}$ at fixed mode period. Since $|v_h|$ does
not suffer from this visibility reduction, velocity variations may be observable 
in stars that are cooler than those at the red edge of the instability strip.  

\begin{figure*}[t]
\centerline{\psfig{figure=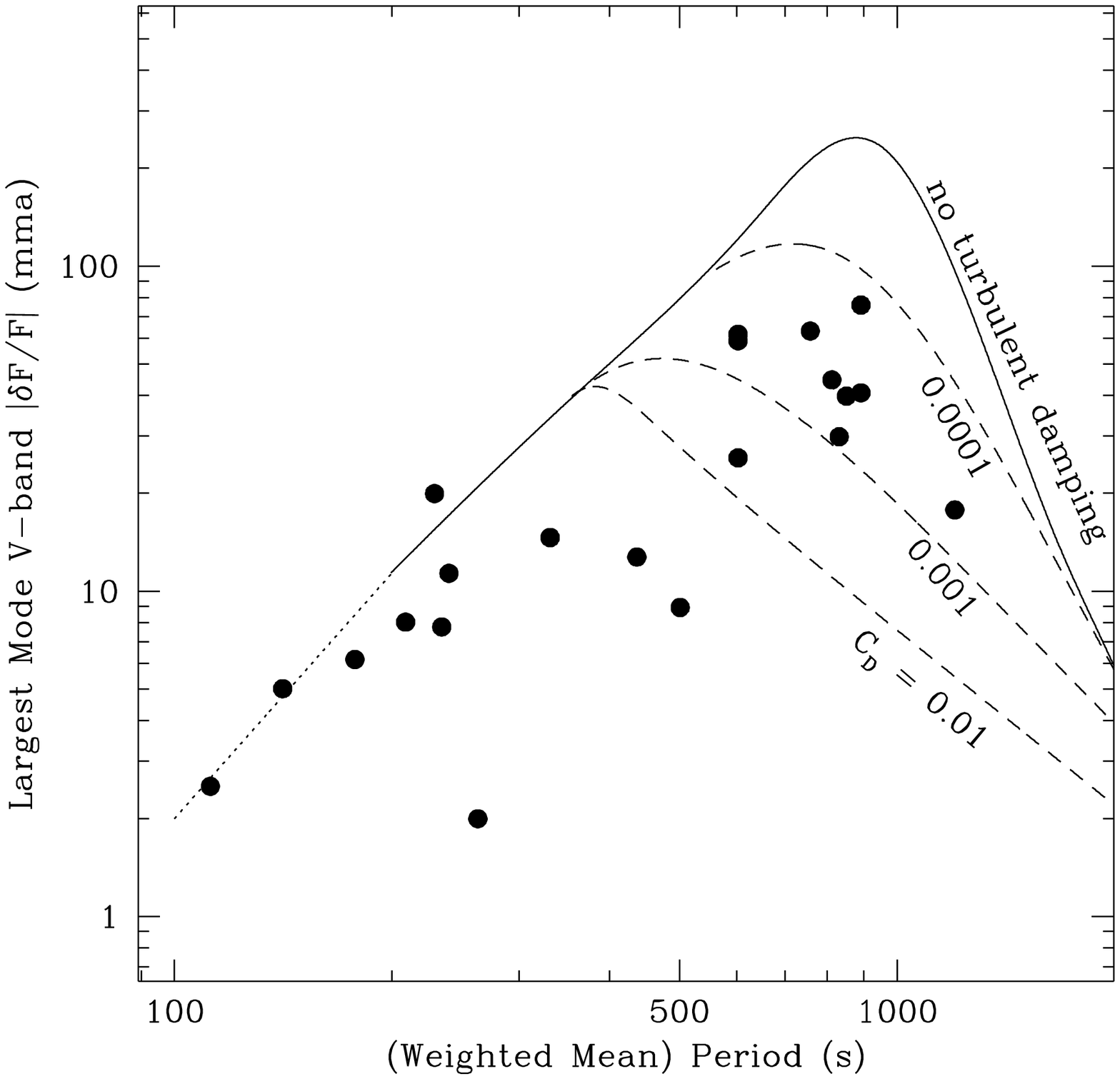,width=0.5\hsize}}
\caption[]{Comparison between theoretical parametric amplitudes and
those summarized by Clemens (\cite{amp-clemens93}) from observations
of two dozen DA variables. Each star is represented by a filled circle
plotted such that the ordinate gives the photometric amplitude of the
largest mode and the abscissa the mean period weighted by the
photometric amplitudes of all its detected modes. To compare theory
and observation, we single out the largest amplitude $\ell = 1$ mode
for each of the three models used in Figure \ref{fig:ampli-final}.
The solid line is an interpolation between these largest amplitude
modes and their periods. The amplitudes are reduced by a factor of
$2.5$ to approximately account for the reduction in visible relative
to bolometric flux variations for DA variables. The run of observed
mode amplitude as a function of mode period is mimicked by the solid
line. However, the extrapolation to short period shown by the dotted
line is not reliable, because in this regime mode amplitudes have
large star-to-star variations associated with different values of
frequency mismatches. The drop-off in amplitude at long period ($P >
10^3 \s$) results from a combination of lower daughter mode damping
rates and a reduction in the visibility of the parent mode due to
increasing $\omega\tau_c$.  Turbulent dissipation arising from the
shear layer below the convection zone (\S \ref{subsec:nl-turbulent})
further suppresses the amplitudes that long period modes can
attain. This is illustrated by the dashed curves for different choices
of the drag coefficient, $C_D$.}
\label{fig:ampli-clemens}
\end{figure*}

\begin{figure*}[t]
\centerline{\psfig{figure=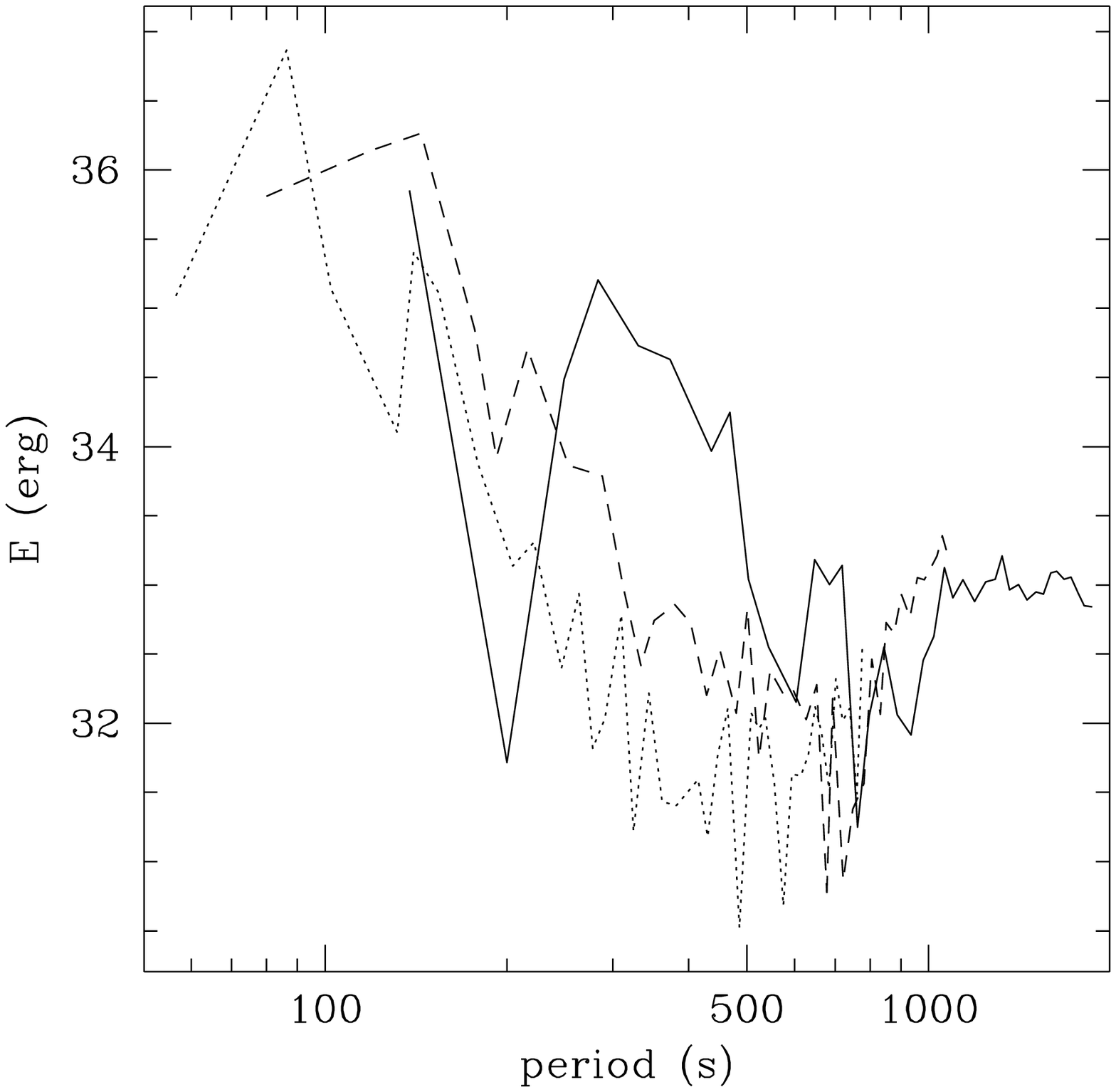,width=0.4\hsize}}
\caption[]{Energy versus period for overstable modes in the coolest
model considered in Figure \ref{fig:ampli-final}. Solid, dashed and
dotted lines represent $\ell = 1$, $2$ and $3$ modes. Mode energy
tends to decrease with increasing period and with increasing
$\ell$. Deviations from this trend for the $\ell=1$ modes are
explained as follows. The narrow dip near $200 \s$ is due to a
fortuitiously good frequency resonance involving the $n=2$ mode, and
the wider dip around $700 \s$ is associated with neutrally stable
daughter modes. }
\label{fig:mode-energy}
\end{figure*}

Figure \ref{fig:ampli-clemens} reproduces a summary of observational
data on mode amplitudes from Clemens (\cite{amp-clemens95}).  Each
star is represented by a point showing the relation between the V-band
photometric amplitude in its largest mode and the amplitude weighted
mean period of all its observed modes. The latter quantity is a
surrogate for the star's effective temperature in the sense that
longer mean periods correspond to lower effective temperatures
(Clemens \cite{amp-clemens95}, Paper I).  Our theoretical predictions
for the amplitudes of $\ell=1$ modes limited by parametric instability
are shown by a solid line obtained by interpolation from the numerical
results displayed in Figure \ref{fig:ampli-final}.  Two features of
this figure are worthy of comment.

For a few low order overstable modes, $n_p \leq 3$ at $\ell_p=1$, $\delta 
\omega$ is more significant than $\gamma_d$ in determining the best daughter 
pair (Fig. \ref{fig:l2-dependence}). The best daughter pairs for these
modes have $\ell$ values of a few and identities which depend sensitively on
minor differences among stars. Thus we expect amplitudes of short period 
overstable modes to show large star to star variations.  That the 
$n=1,\ell=1$ mode is detected in only about one half of the hot DAVs 
(see Figure 4 of Clemens \cite{amp-clemens95}) is consistent with the expected 
statistical variations in $\delta \omega$.

Overstable $\ell=1$ modes with periods longer than $800 \s$ ($n_p \geq 20$) are
probably not saturated by parametric instability. Their maximum $|\kappa|$ is 
severely reduced below the adiabatic value by the strong nonadiabaticity of 
their daughter modes (\S \ref{subsec:app-nonad}). Other mechanisms that  
contribute towards saturating these modes are discussed in \S 
\ref{sec:nl-discussion}.

\section{DISCUSSION}
\label{sec:nl-discussion}

\subsection{Turbulent Saturation}
\label{subsec:nl-turbulent}

Turbulent convection severely reduces the vertical gradient of the horizontal 
velocity of g-modes in the convection zone. As a result, a shear layer forms
at the boundary between the bottom of the convection zone and the top of the 
radiative interior (Goldreich \& Wu \cite{nl-paperIII}, hereafter Paper III). 
Kelvin-Helmholtz instability of this layer provides a nonlinear dissipation 
mechanism for overstable modes. An overstable mode's 
amplitude cannot grow beyond the value at which nonlinear damping due to the 
Kelvin-Helmholtz instability balances its linear convective driving. 
Equation (44) of Paper III provides an estimate for the value at which this
mechanism saturates the surface amplitude of $(\delta p/p)$; 
\be
\left({\delta p\over p}\right)\sim {0.1\over 
C_D}{[(\omega\tau_c)^2+1]^{1/2}[(\omega\tau_c)^2-1]\over 
\omega\tau_c}{L z_\omega^2\over R z_b}.
\label{eq:na-vis-dpnl}
\ee
Our ignorance of the complicated physics involved in a nonlinear shear
layer is covered by the range of possible values of the dimensionless drag 
coefficient, $C_D$. Terrestrial experiments indicate that $C_D$ falls between
$10^{-3}$ and $10^{-1}$.  The dashed lines in Figure
\ref{fig:ampli-clemens} show the effect of including nonlinear
turbulent damping in addition to parametric instability on limiting
mode amplitudes. Amplitudes of overstable modes saturated by the 
Kelvin-Helmholtz instability are uncertain because $C_D$ is poorly constrained.

\subsection{Granddaughter Modes}
\label{subsec:granddaughters}

Here we answer the following questions. Under what conditions do
daughter modes excite granddaughter modes by parametric
instability?\footnote{In this subsection subscripts $p$, $d$, and $g$
refer to parent, daughter, and granddaughter modes.}  What are the
consequences if they do?  In \S \ref{subsec:para-time} we show how
parametric instability of linearly damped daughter modes maintains the
amplitude of a parent mode close to its equilibrium value. In order to
dispose of the energy they receive from the parent mode, the time averaged 
energies of the daughter modes must be close to their
equilibrium values. This raises a worry. Suppose the daughters are
prevented from reaching their equilibrium amplitudes by parametric
instability of granddaughter modes Then they would not be able to halt the
amplitude growth of the parent mode.

To answer the first of these questions, we calculate the ratio,
denoted by the symbol ${\cal S}$, between the threshold amplitude for
a daughter mode to excite granddaughter modes and its equilibrium
amplitude under parametric excitation by the parent mode. The former
is obtained from equation \refnew{eq:thresh}, and the latter from
equations \refnew{eq:nl-amplisteady2}-\refnew{eq:nl-amplisteady3}. We
make a few simplifying assumptions to streamline the
discussion. Resonances between daughters and granddaughters are taken
as exact; individual members of daughter and
granddaughter pairs are treated as equivalent. Equations
\refnew{eq:nl-kappamax} and \refnew{eq:fastgamma} are combined to
yield
\begin{equation} 
\kappa^2\sim \frac{\gamma_p}{n_p^2L}.
\label{eq:nl-kappagamma}
\end{equation}
It is then straightforward to show that
\be
{\cal S}\approx \frac{\omega_p\omega_d}{\omega_g^2} 
\left(\frac{n_d}{n_p}\right)^2 \left(\frac{\gamma_g^2}{\gamma_d^2+\delta 
\omega^2}\right)\approx
32\left(\frac{\gamma_g^2}{\gamma_d^2+\delta\omega^2}\right).
\label{eq:nl-calSp}
\ee
The factor $32$ is an approximation based on taking $\omega_g/\omega_d = 
\omega_d/\omega_p=1/2$ and $n_p/n_d=1/2$. 

In general ${\cal S}\gg 1$, so the excitation of granddaughter modes
requires the daughter modes to have energies in excess of their
equilibrium values.  However, the equilibrium solution is unstable if
the best daughter pair corresponds to $\gamma_d>|\delta\omega|$, and then
the daughter mode energies episodically rise far above their
equilibrium values. At such times, granddaughter modes may be excited
by parametric instability and consequentially limit the amplitude
growth of the daughter modes. This slows the transfer of energy from
parent to daughter modes, but it does not prevent the daughter
modes from saturating the growth of the parent mode's amplitude at the
level described by equation \refnew{eq:thresh}.

For the few lowest order parent modes, we typically find $|\delta
\omega| \geq \gamma_d$. This may reduce ${\cal S}$ to below unity with
the consequence that the parent mode amplitude may rise above that given
by equation \refnew{eq:thresh}.

\subsection{Additional 3-Mode Interactions}
\label{subsec:threemode}

Parametric instability sets reasonable upper bounds on the photospheric
amplitudes of overstable modes.  In a given star this upper bound rises with
increasing mode period except possibly for the lowest few modes.  However, the
observed amplitude distributions are highly irregular.  This mode selectivity
may arise from 3-mode interactions which involve more than one overstable mode.

We investigate a particular example of this type. It is closely related
to parametric instability, the only difference being that the daughter modes of 
the overstable parent mode are themselves overstable. Acting in
isolation, resonant mode couplings tend to drive mode energies toward
equipartition. They conserve the total energy, ${\dot E}_p+{\dot
E}_{d_1}+{\dot E}_{d_2}=0$, and transfer action according to ${\dot
E}_{d_1}/\omega_{d_1} = {\dot E}_{d_2}/\omega_{d_2}=-{\dot
E}_p/\omega_p$. In this context it is important to note that the
energies of overstable modes limited by parametric instability decline
with increasing mode period (Fig. \ref{fig:mode-energy}). Therefore nonlinear 
interactions transfer energy from the parent mode to its independently excited 
daughters. As shown below, this transfer may severely suppress the parent mode's 
amplitude.

We start from equations 
\refnew{eq:nl-amplieqnfulla}-\refnew{eq:nl-amplieqnfullc}. These may be 
manipulated to yield
\begin{equation}
{d E_p\over dt} =\gamma_p E_p+3\sqrt{2}\omega_p\kappa(E_p E_{d_1} 
E_{d_2})^{1/2}\sin\Phi,
\label{eq:nl-Eone}
\end{equation}
where $\Phi = \theta_{d_1} +\theta_{d_2} - \theta_p$. For $E_p\gg
E_{d_1}$ and $E_{d_2}$, nonlinear interactions transfer energy from the parent 
mode to its daughter modes. In
particular, if we ignore phase changes in the overstable daughter
modes due to their interactions with granddaughter modes, we find that $\Phi$ 
satisfies
\begin{equation}
\frac{d\Phi}{dt}=\delta\omega-\frac{3}{\sqrt{2}}\kappa(E_p E_{d_1} 
E_{d_2})^{1/2}
\left[\left(\frac{\omega_{d_1}}{E_{d_1}}+\frac{\omega_{d_2}}{E_{d_2}}\right)- 
\frac{\omega_p}{E_p}\right]\cos\Phi,
\label{eq:nl-Phi}
\end{equation}
with a stable solution at $\Phi=-\pi/2$ when $\delta \omega = 0$.

We denote the ratio of the nonlinear term to the linear term in equation
\refnew{eq:nl-Eone} by the symbol ${\cal T}$; 
\begin{equation} 
{\cal T}\approx \frac{3\sqrt{2}\omega_p\kappa} {\gamma_p} 
\left(\frac{E_{d_1}E_{d_2}}{E_p}\right)^{1/2}.
\label{eq:nl-calT}
\end{equation}
Using the magnitudes of $E_i$ set by parametric instability of their
respective daughters, and adopting the same approximations made in \S
\ref{subsec:granddaughters}, we arrive at
\begin{equation}
{\cal T}\approx \frac{\omega_p\omega_d}{\omega_g^2}
\left(\frac{n_d}{n_p}\right)^2 \left(\frac{\gamma_g^2}{\gamma_d^2 +
\delta \omega^2}\right)\approx
32\left(\frac{\gamma_g^2}{\gamma_d^2+\delta\omega^2}\right).
\label{eq:nl-calTp}
\end{equation}
Comparing equations \refnew{eq:nl-calSp} and \refnew{eq:nl-calTp} we
see that ${\cal S=T}$. A little thought reveals that this is not a
coincidence.  For ${\cal T} \gg 1$, overstable daughter modes can 
suppress a parent mode's energy
below the value set by parametric instability. We expect this
suppression to be important in cool ZZ Ceti stars whose overstable
modes extend to long periods. It may render their intermediate period
modes invisible.  In a similar manner, the amplitudes of high
frequency overstable modes with $\ell=2$ and $3$ may be heavily
suppressed by interactions with their lowest $\ell$ overstable
daughters.

The irregular amplitude distributions among neighboring modes may be
partially accounted for by this type of resonance. Mode variability
may also play a role. We explore this in the next subsection.

\subsection{Mode Variability}
\label{subsec:variability}

Excited g-modes in ZZ Ceti stars exhibit substantial temporal variations. 
Parametric instability may at least partially account for these variations.

When $|\delta \omega| < \gamma_d$, parametric instability gives rise to
limit cycles in which the amplitudes and phases of parent and daughter modes 
vary on time scales as short as $\gamma^{-1}_d$. Stable daughter modes may 
briefly attain visible amplitudes. Temporal amplitude variations may contribute 
to the irregular mode amplitude distribution seen in individual stars.

Phase variations of a parent mode obey the equation
\begin{equation}
{{d\theta_p}\over{dt}} = \omega_p - {3\over{\sqrt{2}}} \omega_p \kappa
{{|A_{d_1}| |A_{d_2}|}\over{|A_p|}} \cos \Phi.
\label{eq:phaseevol}
\end{equation}
At the equilibrium given by equations
\refnew{eq:nl-amplisteady1}-\refnew{eq:nl-amplisteadyphi}, the parent
mode's frequency is displaced from its unperturbed value such that
\begin{equation}
\omega_p^\prime = {{d\theta_p}\over{dt}} = \omega_p +
{{\delta \omega \gamma_p}\over{\gamma_{d_1} + \gamma_{d_2} - \gamma_p}}
\sqrt{1+\left({{2\delta\omega}\over{\gamma_{d_1} + \gamma_{d_2} - 
\gamma_p}}\right)^2}.
\label{eq:equilidomega}
\end{equation} 
This constant frequency shift is of order $10^{-9} s^{-1}$ for the
$n=1$, $\ell=1$ mode and of order $10^{-7} s^{-1}$ for the $n=2$, $\ell=2$ mode. 
Frequency shifts in higher order overstable modes which are involved in limit 
cycles are predicted to be larger and time variable. During brief intervals of 
length $\sim \gamma^{-1}_d$, when the 
daughter mode energies are comparable to that of the parent mode,
$|\omega_p^\prime - \omega_p| \sim \gamma_d$, which is of order a
few times $10^{-5} s^{-1}$, or a few $\mu Hz$ in angular
frequency. These shifts might account for the time-varying rotational
splittings reported by Kleinman \etal (\cite{kleinmanZZPsc}) provided
different $m$ components of the overstable modes are involved in
different limit cycles.

\subsection{Miscellany}
\label{subsec:trapping}

We briefly comment on two relevant issues.

Gravitational settling produces chemically pure layers between which
modes can be partially trapped. Modes that are trapped in the hydrogen
layer have lower mode masses, and therefore higher growth rates and larger
maximum values of $|\kappa|$ than untrapped modes of similar frequency. 
This implies lower threshold energies for parametric instability.
Nevertheless, trapping does not affect predicted photospheric
amplitudes of $\delta p/p$, and hence $v_h$ and $\delta F/F$, since
these are proportional to $A$ divided by the square root of the mode
mass.

In circumstances of small rotational splitting, the simple limit-cycles depicted 
in Figure \ref{fig:nl-detail} are unlikely to be realistic. In such cases, 
different $m$ components of an overstable parent mode share some common daughter 
modes. This leads to more complex dynamics. 

\begin{acknowledgements}

We are indebted to Bradley for supplying us with models of DA white
dwarfs.  Financial support for this research was provided by NSF grant
94-14232.

\end{acknowledgements}

\begin{appendix}

\section{Three-Mode Coupling Coefficient}
\label{sec:app-kappa}

We reproduce the expression for the three-mode coupling coefficient
presented in equation \refnew{eq:app-kappaoriginal} of \S \ref{subsec:nl-kappa}
with one modification:
\begin{equation}
\kappa =  - \int d^3 x p \left\{
{{\Gamma_1 (\Gamma_1 -2)}\over 6} (\divxi)^3
+ {1\over 2} \Gamma_1 (\divxi) \xi^i_{;j} \xi^j_{;i} 
+ {\rm Det}|\xi^i_{;j}|
 \right\},
\label{eq:app-kappamodify}
\end{equation}
where
\begin{equation}
{\rm Det}|\xi^i_{;j}| =  {1\over 6} (\divxi)^3
-{1\over 2} (\divxi) \xi^i_{;j} \xi^j_{;i} 
+ {1\over 3} \xi^i_{;j} \xi^j_{;k} \xi^k_{;i}.
\label{eq:factorJ3}
\end{equation}
Because of strong cancellations among its largest terms, this form is not 
well-suited for numerical evaluation. We derive a new expression which does not
suffer from this defect. Then we estimate the size of $\kappa$ and deduce its 
dependences upon the properties of the three-modes.

\subsection{Simplification}
\label{subsec:app-long}

In spherical coordinates $(r,\theta,\phi)$, the components of the displacement 
vector may be written as
\begin{equation}
{\boldxi} = \xi^i \boldvarepsilon_i = \left[ \xi_r(r), 
{{\xi_h(r)}\over r}
{\partial\over{\partial\theta}}, {{\xi_h(r)}\over r}
 {1\over{\sin^2{\theta}}}{\partial\over{\partial \phi}}\right]
 Y_{{\ell}m}(\theta,\phi),
\label{eq:kappa-xifull2}
\end{equation}
where $Y_{\ell m}$ is a spherical harmonic function, and the 
$\boldvarepsilon_i$ are covariant basis vectors. 

The angular integrations in equation \refnew{eq:app-kappamodify} are
done analytically. The following definitions and properties prove
useful:
\begin{mathletters}
\begin{eqnarray}
T & \equiv & \int_0^{2\pi} d\phi \int_0^{\pi} d\theta \, \sin\theta  Y_a Y_b Y_c 
\equiv
< Y_a Y_b Y_c > , 
\label{eq:Tdefine} \\
F_a & \equiv & < Y_a \nabla_\alpha Y_b \nabla^\alpha Y_c > 
= \left( \Lambda_b^2 +\Lambda_c^2 - \Lambda_a^2\right){T\over 2}  ,
\label{eq:Fdefine}
\\
G_a & \equiv &  < \nabla_\alpha \nabla_\beta Y_a
 \nabla^\alpha Y_b \nabla^\beta Y_c> \equiv 
\left[ \Lambda_a^4 - ( \Lambda_b^2 - \Lambda_c^2 )^2 \right ]{T\over 4} ,
\label{eq:Gdefine} \\
S & \equiv & < \nabla_\alpha (\nabla^\alpha Y_a \nabla_\beta Y_b
\nabla^\beta Y_c)> = G_a + G_b + G_c \nonumber \\
& = & {1\over 2} (\Lambda_a^2 F_a +
\Lambda_b^2 F_b + \Lambda_c^2 F_c) = G_a + \Lambda_a^2 F_a,
\label{eq:Sdefine} \\
V_a & \equiv & < Y_a \nabla_\alpha \nabla^\beta Y_b \nabla_\beta \nabla^\alpha 
Y_c> 
= \Lambda_b^2 \Lambda_c^2 T - F_a - S,
\label{eq:defineV} \\
\Lambda_a^2 T & = & F_b + F_c,
\label{eq:equalF} \\
\Lambda_a^2 F_a & = & G_b + G_c.
\label{eq:equalG}
\end{eqnarray}
\end{mathletters}
\noindent Here, $\nabla_\alpha$ is the covariant derivative on a spherical
surface; $\alpha$ can be either $\theta$ or $\phi$. Each angular integration is 
proportional to $T$ which contains all $m$ dependences and is of order unity 
independent of the $\ell$ values of the participating modes. The paramter 
$\Lambda^2 \equiv \ell(\ell+1)$. Subscripts $a$, $b$ and $c$ denote different 
modes. The angular selection rules are simply, ${\ell}_c \in [|{\ell}_a -
{\ell}_b|, {\ell}_a + {\ell}_b]$, Mod$[({\ell}_c + {\ell}_b +
{\ell}_c), 2] = 0$ and $m_a +m_b + m_c = 0$. 

Angular integration of ${\rm Det}|\xi^i_{;j}|$ yields
\begin{eqnarray}
\oint d\Omega\, {\rm Det}|\xi^i_{;j}|  & = & 
{1\over r^2} 
\left[ T \xi^b_r \xi^c_r - \Lambda_c^2  T \xi^b_r \xi^c_h 
+ {1\over 2}{(\Lambda_b^2 \Lambda_c^2 T -  V_a)}
\xi^b_h \xi^c_h \right] {{d\xi^a_r}\over {dr}}  \nonumber\\
& & + {1\over r^2} \left[ S (\xi^b_r - \xi^b_h) \xi^c_h 
 - F_b \xi^b_r (\xi^c_r - \xi^c_h)\right] {{d\xi^a_h}\over{dr}} .
\label{eq:firstk3}
\end{eqnarray}
Symmetrizing this expression with respect to modes $b$ and $c$,
employing equations \refnew{eq:Sdefine}-\refnew{eq:equalG}, we get
\begin{equation}
\oint d\Omega\,  {\rm Det}|\xi^i_{;j}| =  
- {S \over{3 r^2}}{d\over{dr}}\left(\xi^a_h \xi^b_h \xi^c_h\right) 
+ {{(F_a + S)}\over{2 r^2}} {d\over{dr}}\left(\xi^a_r \xi^b_h \xi^c_h\right) 
- {{\Lambda_a^2 T}\over{2 r^2}} {d\over{dr}}\left(\xi^a_h \xi^b_r \xi^c_r\right) 
+ {T\over {3r^2}} {d\over{dr}}\left(\xi^a_r \xi^b_r \xi^c_r\right).
\label{eq:secondk3}
\end{equation}
For gravity-modes, $|\xi_h|\gg |\xi_r|$. Thus the four terms in this
expression decrease in size from left to right. 

In a similar manner we arrive at
\begin{equation}
\oint d\Omega\, (\divxi)^3 = T(\divxi^a) (\divxi^b) (\divxi^c),
\label{eq:firstk1}
\end{equation}
and
\begin{eqnarray} 
\oint d\Omega\, (\divxi) \xi^i_{;j} \xi^j_{;i} & = & - {{F_a}\over r}
(\divxi^a) {d\over{dr}}\left(\xi^b_h \xi^c_h\right) + {{2 F_a}\over r}
(\divxi^a) \xi^b_r {{d\xi^c_h}\over{dr}} + T (\divxi^a) {{d\xi^b_r}\over{dr}}
{{d\xi^c_r}\over{dr}} \nonumber \\
 & & + {{V_a}\over{r^2}}(\divxi^a) \xi^b_h \xi^c_h  
 -{{2T \Lambda_c^2}\over{r^2}} (\divxi^a) \xi^b_r \xi^c_h
 + {{2 T}\over{r^2}} (\divxi^a) \xi^b_r \xi^c_r.
\label{eq:firstk2}
\end{eqnarray}
The magnitudes of the terms in the above expression decrease from left to right 
except that terms 2-4 are of comparable size. 

Having disposed of the angular dependences in $\kappa$, we turn to the radial 
integrations.  The following relations prove helpful in this context:
\begin{eqnarray}
{{d\xi_r}\over{dr}} & = & (\divxi) + {{\Lambda^2}\over r}\xi_h - {2\over 
r}\xi_r,
\label{eq:radial1} \\
{d\over{dr}} \left[\Gamma_1 p (\divxi)\right]&  =&  {{\Lambda^2\rho g }\over
r} \xi_h - \left(\omega^2 + {{2g}\over r} - {{dg}\over{dr}}\right)
\rho \xi_r,
\label{eq:radial2} \\
\Gamma_1 p (\divxi) & = & \rho g \xi_r - r \rho \omega^2 \xi_h.
\label{eq:radial3}
\end{eqnarray}
The first equation is the definition of divergence, and the second and third are
the radial and horizontal components of the equation of motion written in terms 
of the Lagrangian displacement. 

Curvature terms arise because the directions of the basis vectors depend upon 
position. For instance,
\begin{equation}
\xi^r_{;\theta} = (\xi_r - \xi_h) {{\partial Y_{\ell m}}\over{\partial \theta}},
\label{eq:covexample}
\end{equation}
where $\xi_h$ appears due to curvature in the coordinate system. The
largest terms in equations \refnew{eq:secondk3} and
\refnew{eq:firstk2} are curvature terms. However, their radial
integrals cancel leaving a much smaller net contribution. Direct
numerical evaluation of equation \refnew{eq:app-kappamodify} leads to
unreliable results, so it is important to carry out this
cancellation analytically. Accordingly, we integrate each term by parts
and apply equations \refnew{eq:Sdefine} and \refnew{eq:radial1} to
obtain
\begin{eqnarray}
& &  \int_0^R dr\,\left[ {S p\over 3}{d\over{dr}}\left(\xi^a_h \xi^b_h \xi^c_h
\right) + {{F_a r \Gamma_1 p}\over 2} (\divxi^a)
{d\over{dr}}\left(\xi^b_h \xi^c_h\right)\right] \nonumber \\
& & = - \int_0^R dr\, {{F_a\Gamma_1 p}\over 2} (\divxi^a) \xi^b_h \xi^c_h + 
\int_0^R dr\, {{F_a r\rho}\over 2} \left(\omega_a^2 + {{2g}\over r} - 
{{dg}\over{dr}}\right)
\xi^a_r \xi^b_h \xi^c_h.
\label{eq:cancel}
\end{eqnarray}
This step leads to
\begin{eqnarray}
\kappa & = & \int_0^R dr\, \left[
-{T r^2 {\Gamma_1 (\Gamma_1 - 2) p}\over{6}} (\divxi^a) (\divxi^b) (\divxi^c) 
- {T r^2 {\Gamma_1 p}\over 2} (\divxi^a) {{d\xi^b_r}\over{dr}} 
{{d\xi^c_r}\over{dr}} 
\right.
\nonumber \\
& & \left. \hskip-0.8cm
-  F_a r\Gamma_1 p (\divxi^a) \xi^b_r {{d\xi^c_h}\over{dr}}
- {(\Lambda_b^2 \Lambda_c^2 T - S){\Gamma_1 p}\over{2}} (\divxi^a) \xi^b_h 
\xi^c_h
-{{(S- F_a)\rho g }\over{2}} \xi^a_r \xi^b_h \xi^c_h 
- {{F_a r \rho}\over{2}}{{dg}\over{dr}} \xi^a_r \xi^b_h \xi^c_h   
\right. \nonumber \\
& & \left. \hskip-0.8cm
+ \Lambda_a^2 T \rho g \xi^a_h \xi^b_r \xi^c_r 
+ \Lambda_c^2 T \Gamma_1 p(\divxi^a) \xi^b_r \xi^c_h 
- T p {{d\xi^a_r}\over{dr}} \xi^b_r \xi^c_r 
- T \Gamma_1 p (\divxi^a) \xi^b_r \xi^c_r
+ {{\omega_a^2 F_a r \rho }\over{2}} \xi^a_r \xi^b_h \xi^c_h 
\right].
\label{eq:workingkappa}
\end{eqnarray}

Next we systematically eliminate radial derivatives of the
displacement vector from the expression for $\kappa$.  We integrate by
parts to dispose of $d\xi_h/dr$ and substitute for $d\xi_r/dr$ using
equation \refnew{eq:radial1}.  With the aid of equations
\refnew{eq:Sdefine}-\refnew{eq:equalF}, and using equation
\refnew{eq:radial3} to make the expression symmetric with respect to
indexes $b$ and $c$, we arrive at our final working expression for
$\kappa$,
\begin{eqnarray}
\kappa & = & \int_0^R dr\, \left\{ 
- {Tr^2{\Gamma_1 (\Gamma_1 + 1) p}\over 6}  (\divxi^a) (\divxi^b) (\divxi^c)
+ {{\omega_a^2 G_a r\rho }\over 2} \xi^a_h \xi^b_h \xi^c_h 
\right. \nonumber \\ & & \left. \hskip-0.8cm 
+ {{F_a\rho}\over 2}\left(g-r{{dg}\over{dr}}\right)\xi^a_r \xi^b_h \xi^c_h 
- {\Lambda_a^2 T r{\Gamma_1 p}\over 2} \xi^a_h (\divxi^b) (\divxi^c)
- T(3 \Gamma_1 + 1)p(\divxi^a) \xi^b_r \xi^c_r \right. \nonumber \\ & & \left.   
\hskip-0.8cm + 2 T r\Gamma_1 p\xi^a_r (\divxi^b) (\divxi^c) 
+ {{2T p}\over r} \xi^a_r \xi^b_r \xi^c_r
+ {1\over 2}\left[\left(\omega_a^2-3\omega_b^2-3\omega_c^2\right)F_a 
-(2\omega_b^2F_b +2\omega_c^2F_c)\right] r\rho\xi^a_r \xi^b_h \xi^c_h
\right. \nonumber \\ & & \left. \hskip-0.8cm 
+ {1\over 2}\left[\Lambda_a^2 T\left(5\rho g+\rho r{dg\over{dr}}-{2p\over 
r}\right)
-\rho \left(\omega_b^2F_b+\omega_c^2F_c\right)r\right]\xi^a_h \xi^b_r \xi^c_r 
\right\}
\label{eq:working2}
\end{eqnarray}
All permutations of the 3 modes are to be included when evaluating this 
expression. However, the largest contribution from each term comes when 
$b$ and $c$ are radially similar daughter modes. For high order gravity-modes, 
the first five terms are of comparable size and much larger than the remaining 
four. Numerical evaluation of $\kappa$ using equation \refnew{eq:working2} does 
not suffer from the errors arising from large cancellations and numerical 
differentiation that plague attempts using equation \refnew{eq:app-kappamodify}.

\subsection{Order-of-Magnitude}
\label{subsec:app-orders}

Here we estimate the maximum value that $\kappa$ can attain for parametric 
resonances involving a given parent mode. In so doing, we apply results 
derived in Papers I and II. These include: (1) the scaling relations
\begin{equation}
|\divxi|  \sim {{\Lambda^2}\over R} \xi_h \sim {{\xi_r}\over{z_\omega}}\approx 
k_h \xi_h \sim {1\over (n\tau_\omega L)^{1/2}},
\label{eq:evanescent}
\end{equation}
in the evanescent region $z< z_\omega$, and
\begin{equation}
|\divxi| \sim \left({{z_\omega}\over z}\right)^{1\over 2} 
{{\Lambda^2}\over R} \xi_h \sim  {{\xi_r}\over{z}},
\label{eq:propagating}
\end{equation}
in the propagating cavity $z > z_\omega$; 
(2) the fact that regions between consecutive radial nodes contribute
equally to the following normalization integral
\begin{equation}
{\omega^2\over 2} \int_0^R dr\, r^2 \,\rho\left(\Lambda^2 \xi_h^2 
+\xi_r^2\right)=1.
\label{eq:normalization}
\end{equation}
As $\xi_r^2\ll \xi_h^2$ for g-modes, we have
\begin{equation}
\omega^2 R^2 \int_0^z dz \, \rho \Lambda^2 \xi^2_h \sim
{{n^\prime}\over{n}} \sim {1\over{n}}{\int_0^z dz\,
k_z} \sim {1\over n} \left({z\over{z_\omega}}\right)^{1\over 2}.
\end{equation}
Here $n^\prime$ is the number of radial nodes above depth $z$ and $n$
the total number in the mode.

Equation \refnew{eq:working2} yields maximal values for
$\kappa$ when mode $a$ is taken to be the parent mode and modes $b$
and $c$ to be two radially similar daughter modes. Most of the
contribution to the radial integral comes from the region above
$z_{\omega_a}$: for parametric resonance, $z_{\omega_a}$ is much
greater than $z_{\omega_b}\approx z_{\omega_c}$; the decay and rapid
oscillation of the parent mode's eigenfunction renders insignificant
contribution from greater depths.  Thus we can take the integrals in
equation \refnew{eq:working2} to run from $z=0$ to $z=z_{\omega_a}$
and pull out the parent mode eigenfunctions since they are
approximately constant for $z<z_{\omega_a}$.  This procedure reduces
each of the leading terms in $\kappa$ and thus their sum to 
\begin{equation}
\kappa
\sim \left(\frac{z_{\omega_p}}{z_{\omega_d}}\right)^{1/2}{1\over
n_d(n_p\tau_{\omega_p} L)^{1/2}}.
\label{eq:prelimkappa}
\end{equation}
But $z_{\omega_p}/z_{\omega_d}\sim \Lambda^2_d/\Lambda^2_p\sim n^2_d/n^2_p$ 
which
leads to
\be
\kappa\sim
{1\over (n_p^3\tau_{\omega_p} L)^{1/2}}.
\label{eq:finalkappa}
\ee
The above equation establishes that the maximum value of $\kappa$ rises steeply 
with increasing radial order of the parent mode and is independent of the radial 
orders and spherical degrees of the daughter modes. 

\section{STRONGLY NONADIABATIC DAUGHTER MODES}
\label{sec:nl-highl}

Strong nonadiabaticity occurs wherever
\be
\frac{\omega\tau_{\rm th}}{(k_z z)^2}\lesssim 1.
\label{eq:app-strongna}
\ee
We refer to the depth above which this inequality applies as $z_{\rm na}$. To 
derive a simple analytic scaling relation for $z_{\rm na}$, we
make use of the approximations $k_z\sim (z z_{\omega})^{-1/2}$ and $\tau_{\rm 
th}/\tau_b\sim (z/z_b)^{q+2}$. Here $z_b$ and $\tau_b$ are the depth and thermal
relaxation time at the bottom of the convection zone, and $\rho\propto z^q$ 
with $q\approx 3.5$ provides a fit to the density structure in the upper portion 
of the radiative interior. It then follows that
\be
{z_{\rm na}\over z_b}\sim \left({1\over \omega\tau_b}{z_b\over 
z_\omega}\right)^{{1\over(q+1)}},
\label{eq:app-zna}
\ee
Moreover,
\be
\left(k_zz\right)_{\rm na}\sim \left[{1\over(\omega\tau_b)}\left({z_b\over 
z_\omega}\right)^{q+2}\right]^{{1\over 2(q+1)}}.
\label{eq:app-kzzna}
\ee
By reducing the effective buoyancy, strong nonadiabaticity lowers the effective 
lid of a g-mode's cavity to $z_{\rm na}$. Consequences of this fact are
explored in the following subsections.

\subsection{Damping Rates of Strongly Nonadiabatic Modes}
\label{subsec:app-gamma}

As shown in Paper II, the energy dissipation rate for a strongly nonadiabatic
mode may be written as
\be
\gamma\approx \frac{\omega}{\pi n}\ln{\cal R}^{-1},
\label{eq:app-omegai}
\ee
where ${\cal R}$ denotes the coefficient of amplitude reflection at the top of 
the mode's cavity. 

\begin{figure*}[t]
\centerline{\psfig{figure=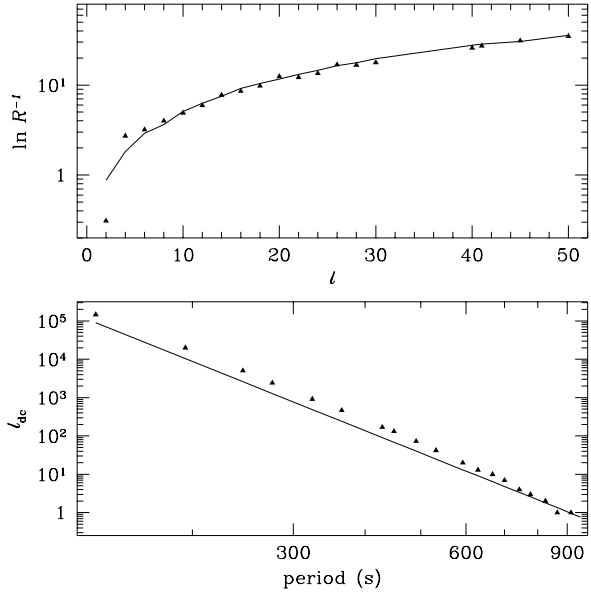,width=0.5\hsize}}
\caption[]{Strongly nonadiabatic modes in the hottest model considered
in Figure \ref{fig:ampli-omegai}. The top figure exhibits the
dependence of the reflection coefficient, $\cal R$, on $\ell$ for
modes with periods close to $550 \s$. The numerical result for the
reflection coefficient is obtained from numerical computations of
$\gamma$ and is depicted by filled triangles. The underlying solid
line is the theoretical estimate (eq. [\ref{eq:app-calRp}]). Computing
$\gamma$ requires care since ${\cal R}\approx 10^{-16}$ for $\ell =
50$. This is of order the machine accuracy for double precision
numbers.  The bottom panel shows the angular degree at which
decoupling occurs, $\ell_{\rm dc}$, as a function of period for
$\ell=1$ parent modes.  Triangles denote numerical values and the
solid line the analytical scaling relation given by equation \refnew
{eq:app-dc}. We find that $\ell_{\rm dc}$ is almost independent of the
star's effective temperature.}
\label{fig:omegai-split}
\end{figure*}

To derive an approximate relation for ${\cal R}$, we note that the real and 
imaginary parts of $k_z$, $k_{zr}$ and $k_{zi}$, satisfy
\be
\frac{|k_{zi}|}{|k_{zr}|}\sim \frac{(k_{zr} z)^2}{\omega\tau_{\rm th}}, 
\label{eq:app-kikr}
\ee
provided $|k_{zi}|/|k_{zr}|\lesssim 1$. Thus
\be
\ln{\cal R}^{-1}\sim \int^\infty_{z_{\rm na}} dz\, |k_{zi}|.
\label{eq:app-calR}
\ee
Evaluating this integral with the aid of equations \refnew{eq:app-zna} and
\refnew{eq:app-kzzna}, we obtain
\be
\ln{\cal R}^{-1}\sim \left[{1\over(\omega\tau_b)}\left({z_b\over 
z_\omega}\right)^{q+2}\right]^{{1\over 2(q+1)}} \propto
\frac{\ell^{(q+2)/(q+1)}}{\omega^{(2q+5)/(2q+2)}}.
\label{eq:app-calRp}
\end{equation} 
Numerical results for $\ln{\cal R}^{-1}$ plotted in the upper panel of
Figure \ref{fig:omegai-split} confirm this relation.

Because the maximum value of $\kappa$ for parametric resonance is
independent of the spherical degrees of the daughter modes, the
dependence of $\gamma$ on $\ell$ at fixed $\omega$ is of great
significance. Equations \refnew{eq:app-omegai} and
\refnew{eq:app-calRp}, together with the relation $n\propto \ell$ at
fixed $\omega$, establish that $\gamma\propto \ell^{1/(q+1)}$. Since
$\gamma$ increases with $\ell$ at fixed $\omega$, the most important
daughter pairs are those with the smallest $\ell$ values subject to
the constraint $\gamma_d>\delta\omega$.

\subsection{Reduction of the Coupling Coefficient by Strong Nonadiabaticity}
\label{subsec:app-nonad}

The maximum adiabatic coupling coefficient between a parent mode and a
pair of daughter modes is $\kappa_{\rm max}\sim (n_p^3 \tau_{\omega_p}
L)^{-1/2}$ (eq.  [\ref{eq:nl-kappamax}]).  Its major contribution
comes from the region above $z_{\omega_p}$.  The factor $(n_p
\tau_{\omega_p} L)^{-1/2}$ is the surface value of the normalized
eigenfunction for $[\ell(\ell+1)]^{1/2}\xi_p/R$.  The extra factor
$n_p^{-1}$ is the fraction of each daughter mode's nodes that lie
above $z_{\omega_p}$.  The coupling coefficient is reduced compared to
equation \refnew{eq:nl-kappamax} for daughter modes that are strongly
nonadiabatic in the region above $z_{\omega_p}$. 

Equation \refnew{eq:app-strongna} indicates that nonadiabaticity
increases with increasing $\ell$ at fixed $z$ and $\omega$; $k_z\sim
(z z_\omega)^{-1/2}$ and $z_\omega\sim \omega^2 R^2/g\ell(\ell+1)$, so
$k_z\propto \ell$. We define the angular degree of decoupling for a
given parent mode, $\ell_{\rm dc}$, as the smallest spherical degree
at which its daughter modes are strongly nonadiabatic all the way down
to the top of the parent mode's cavity; that is, $z_{\rm na_d}\approx
z_{\omega_p}$ at $\ell_{\rm dc}$. Using equation
\refnew{eq:app-zna}, we find 
\begin{equation} 
\ell_{\rm dc}\sim
\left[\left(\frac{\omega^2 R^2}{gz_b}\right)^{q+2}
\frac{\omega\tau_b}{(\ell_p(\ell_p+1))^{q+1}}\right]^{1/2}.
\label{eq:app-dc}
\end{equation}
Numerical results displayed in the lower panel of Figure
\ref{fig:omegai-split} are well-represented by this scaling
relation. We find that $\ell_{\rm dc}$ is relatively independent of stellar
effective temperature but decays steeply with mode period. By $n_p =
20$, we find $\ell_{\rm dc} \leq 1$.

It is plausible that at $\ell_d = \ell_{\rm dc}$, $\kappa_{\rm max}$
is reduced by a factor $n_p/n_d$ below its adiabatic value
(eq. [\ref{eq:nl-kappamax}]) because the effective lids of the
daughter modes' cavities are lowered to $z_{\omega_p}$. Since
$n_p/n_d\approx \ell_p/2\ell_d$, this is a large reduction for 
$\ell_{\rm dc}\gg \ell_p$. An even more severe reduction of
$\kappa_{\rm max}$ is expected for $\ell_d>\ell_{\rm dc}$.

\subsection{Parametric Instability for Traveling Waves
\label{subsec:app-travel}}

In the limit of strong nonadiabaticity, daughter modes are more
appropriately described as traveling waves than as standing
waves. Thus it behooves us to investigate the parametric instability
of traveling waves. Here we demonstrate that the instability criterion
for traveling waves is equivalent to that for standing waves
(eq. [\ref{eq:thresh}]).

Nonlinear interactions between parent and daughter modes are localized
within an interaction region above $z_{\omega_p}$. Let us assume that
$z_{\rm na_d}\ll z_{\omega_p}$. Then, in most of the interaction
region daughter wave packets may be represented as linear
superpositions of adiabatic modes.  Propagating at their group
velocity, the daughter wave packets pass through the interaction
region in a time interval 
\begin{equation} 
\Delta T = \int_0^{z_{\omega_p}}
{dz\over v_{gz}} \sim {1\over n_p} {\pi n_d\over \omega_d}.
\label{eq:ampli-widthwave}
\end{equation}

Three significant relations involving $n_p$ are worth noting: 1)
$n_p^{-1}$ is the fraction of each daughter mode's nodes that lie
above $z_{\omega_p}$, so $2\Delta T$ is a fraction $n_p^{-1}$ of the
time each daughter wave packet takes to make a round trip across its
cavity; 2) approximately $n_p$ daughter modes reside within the
frequency interval $\pi/\Delta T$, and their relative phases change by
less than $\pi$ as each daughter wave packet crosses the interaction
region; 3) maximal $\kappa$ occurs inside an interval of width
$|n_{d_1} - n_{d_2}| \lesssim n_p$.

Nonlinear interactions between parent and daughter waves within the
interaction region are described by equations
\refnew{eq:nl-amplieqnfulla}-\refnew{eq:nl-amplieqnfullc} with two
modifications of the equations governing the time evolution of the
daughter modes. The linear damping term is negligible for $z\gg z_{\rm
na}$, and the nonlinear term must be multiplied by a factor $n_p$. The
latter accounts for the number of modes which couple coherently to
each daughter mode during the interaction time $\Delta T$. During two
passes through the interaction region, the amplitudes of the daughter
wave packets grow by a factor $e^G$, where the gain, ${\cal G}$, is
given by 
\begin{equation} 
{\cal G} = {{2 \Delta T}\over |A_d|}{{d|A_d|}\over{dt}}
\approx 3\sqrt{2} n_d |\kappa| |A_p|.
\label{eq:nl-GA1}
\end{equation}
For parametric instability to occur, 
\begin{equation}
{\cal G}>\ln{\cal R}^{-1}.
\label{eq:nl-GRrelation}
\end{equation} 
Combining the relation between ${\cal R}$ and $\gamma$ given by
equation \refnew{eq:app-omegai} with equation
\refnew{eq:nl-GRrelation}, the threshold condition for parametric
instability of traveling waves becomes
\begin{equation}
|A_p|>\frac{\gamma_d}{3\sqrt{2}\omega_d|\kappa|}.
\label{eq:threshtravel}
\end{equation} 
The above condition is equivalent to equation \refnew{eq:thresh}
in the limit that $\gamma_d\gg \delta\omega$. Thus the threshold
condition for parametric instability of traveling waves reduces to a
limiting case of the threshold condition for parametric instability of
standing waves.

\end{appendix}


\begin{thebibliography}{}
\bibitem[1996]{nl-bradley96}Bradley, P. A. 1996, \apj, 468, 350
\bibitem[1990]{nonad-brick90} Brickhill, A. J. 1990, \mnras, 246, 510
\bibitem[1991]{nonad-brick91} Brickhill, A. J. 1991, \mnras, 251, 673
\bibitem[1993]{amp-clemens93} Clemens, J. C. 1993, Baltic Astronomy, 2, 407
\bibitem[1995]{amp-clemens95} Clemens, J. C. 1995, Baltic Astronomy,
4, 142
\bibitem[1982]{nl-dziem82} Dziembowski, W. 1982, Acta Astronomica, 32,
147
\bibitem[1985]{nl-dziem85} Dziembowski, W., \& Krolikowska, M. 1985, Acta
Astron. 35, 5
\bibitem[1998]{nl-paperI} Goldreich, P. \& Wu, Y. 1999, \apj, 511,904 (Paper I)
\bibitem[1999]{nl-paperIII} Goldreich, P. \& Wu, Y. 1999, \apj, 523, 805 (Paper 
III)
\bibitem[1998]{kleinmanZZPsc} Kleinman, S. J., Nather, R. E. Winget,
D. E., \etal 1998, \apj, 495,424
\bibitem[1989]{nl-kumar89} Kumar, P., \& Goldreich, P. 1989, \apj, 342, 558
\bibitem[1996]{nl-kgood96} Kumar, P., \& Goodman, J. 1996, \apj , 466, 946
\bibitem[1976]{nl-landau76} Landau, L. D., \& Lifshitz, E. M. 1976,
Mechanics, Third Edition, (Pergamon Press), 80
\bibitem[1962]{nl-newcomb62} Newcomb, W. A. 1962, Nuclear Fusion: Supplement 
Part 2, Vienna: International Atomic Energy Energy Agency, 451
\bibitem[1979]{nl-vandakurov79} Vandakurov, V. V. 1979, \azh, 56, 749
\bibitem[1980]{nl-wersinger80} Wersinger, J. M., Finn, J. M., \& Ott, E. 1980, 
Physics of Fluids, 23, 1142
\bibitem[1999]{nl-paperII} Wu, Y. \& Goldreich, P. 1999, \apj, 519, 783 (Paper 
II)
\end{thebibliography}
\end{document}